\documentclass[1p,preprint,10pt]{elsarticle}
\pdfoutput=1 
\usepackage{calc}
\usepackage{amsmath}
\usepackage{amssymb}
\usepackage{amsmath}
\usepackage{textcomp}
\usepackage{color}
\usepackage{framed}
\usepackage{ctable}
\usepackage{tabularx}
\usepackage{hyperref}
\usepackage{wrapfig}
\usepackage{xspace}

\hypersetup{colorlinks=true,breaklinks=true,linkcolor=blue,urlcolor=blue,citecolor=blue}
\definecolor{shadecolor}{rgb}{0.9,0.9,0.9}

\journal{Computer Physics Communications}

\usepackage{graphicx}

\usepackage{amssymb}

\usepackage{lineno}

\definecolor{darkblue}{rgb}{0,0,.6}
\definecolor{darkred}{rgb}{.6,0,0}
\definecolor{darkgreen}{rgb}{0,.6,0}
\definecolor{red}{rgb}{.98,0,0}

\def\ssmall{\fontsize{8pt}{2pt}\selectfont}

\usepackage{listings}
\usepackage{caption}

\DeclareCaptionFormat{listing}{\rule{\dimexpr\textwidth+17pt\relax}{0.4pt}\par\vskip1pt#1:#2#3}
\lstnewenvironment{cpplisting}[1][]
  {\lstset{language=C++,numbers=right,#1}}{}

\lstnewenvironment{bashlisting}[1][]
  {\lstset{language=bash,numbers=right,#1}}{}

\lstnewenvironment{outputlisting}[1][]
  {\lstset{language=html,numbers=right,#1}}{}

\newcommand{\code}[1]{\texttt{#1}}

\newcommand{\Boost}{\code{Boost}\xspace}
\newcommand{\MPI}{\code{MPI}\xspace}
\newcommand{\Doxygen}{\code{Doxygen}\xspace}
\newcommand{\CXX}{\code{C++}\xspace}
\newcommand{\Python}{\code{Python}\xspace}
\newcommand{\CMake}{\code{CMake}\xspace}
\newcommand{\GCC}{\code{GCC}\xspace}
\newcommand{\HDF}{\code{HDF5}\xspace}
\newcommand{\Git}{\code{Git}\xspace}
\newcommand{\OpenMP}{\code{OpenMP}\xspace}

\newcounter{bla}

\begin{document}

\begin{frontmatter}



\title{Updated Core Libraries of the ALPS Project}

\newcommand{\ToBeConfirmed}{\textcolor{blue}{ (TBC)}}
\newcommand{\declined}{\textcolor{red}{ (DECLINED!)}}
\author[umich]{A.~Gaenko\corref{author}} \ead{galexv@umich.edu}
\author[umich]{A.~E.~Antipov}  
\author[umich]{G.~Carcassi}
\author[wchesterpenn]{T.~Chen}
\author[umich]{X.~Chen}         
\author[umich]{Q.~Dong}      
\author[eth]{L.~Gamper}                            
\author[sherbrooke]{J.~Gukelberger}                       
\author[utokyo1]{R.~Igarashi}   
\author[umich]{S.~Iskakov}  
\author[eth]{M.~K\"{o}nz}   
\author[umich,mun]{J.~P.~F.~LeBlanc} 
\author[umich,uiuc]{R.~Levy}
\author[spore]{P.~N.~Ma}                          
\author[umich]{J.~E.~Paki}             
\author[saitama]{H.~Shinaoka}
\author[utokyo2,utokyo3]{S.~Todo}             
\author[eth]{M.~Troyer}
\author[umich]{E.~Gull}
\cortext[author]{Corresponding author.}

\address[umich]{University of Michigan, Ann Arbor, Michigan 48109, USA}
\address[wchesterpenn]{West Chester University of Pennsylvania, West Chester, Pennsylvania 19383, USA}
\address[eth]{Theoretische Physik, ETH Zurich, 8093 Zurich, Switzerland}
\address[sherbrooke]{D\'epartement de Physique, Universit\'e de Sherbrooke, Sherbrooke, Qu\'ebec, J1K 2R1, Canada}
\address[utokyo1]{Information Technology Center, the University of Tokyo, Tokyo 113-8658, Japan}
\address[mun]{Department of Physics and Physical Oceanography, Memorial University of Newfoundland, St. John’s, Newfoundland \& Labrador A1B 3X7, Canada}
\address[uiuc]{University of Illinois, Urbana-Champaign, Illinois 61820, USA}
\address[spore]{National University of Singapore, Singapore}
\address[saitama]{Department of Physics, Saitama University, Saitama 338-8570, Japan}
\address[utokyo2]{Department of Physics, University of Tokyo, Tokyo 113-0033, Japan}
\address[utokyo3]{Institute for Solid State Physics, University of Tokyo, Kashiwa 277-8581, Japan}

\begin{abstract}
The open source ALPS (Algorithms and Libraries for Physics Simulations) project provides a collection of physics libraries and applications, with a focus on simulations of lattice models and strongly correlated systems. The libraries provide a convenient set of well-documented and reusable components for developing condensed matter physics simulation code, and the applications strive to make commonly used and proven computational algorithms available to a non-expert community. In this paper we present an updated and refactored version of the core ALPS libraries geared at the computational physics software development community, rewritten with focus on documentation, ease of installation, and software maintainability.
\end{abstract}

\end{frontmatter}



\noindent {\bf PROGRAM SUMMARY}

\begin{small}
\noindent
{\em Program Title:} ALPS Core libraries                                     \\
{\em Project homepage:} http://alpscore.org                \\
{\em Catalogue identifier:} --                                  \\
{\em Journal Reference:} --                                     \\
{\em Operating system:} Unix, Linux, OSX \\
{\em Programming language:} \verb*#C++#\\
{\em Computers:}
  any architecture with suitable compilers including PCs, clusters and
  supercomputers\\
{\em RAM:} Highly problem-dependent \\
{\em Distribution format:} GitHub, downloadable as zip \\
{\em Licensing provisions:} GNU General Public License\\
{\em Classification:}
6.5, 
7.3, %
20 
{\em External routines/libraries:}
  \CMake, 
  \MPI, 
  \Boost, 
  \HDF. \\ 
{\em Nature of problem:}
Need for modern, lightweight, tested and documented libraries covering
the basic requirements of rapid development of efficient physics
simulation codes, especially for modeling strongly correlated electron systems. \\
{\em Solution method:}\\
We present a \CXX open source computational library that
provides a convenient set of components for developing parallel
physics simulation code. The library features a short development
cycle and up-to-date user documentation.
\\
{\em Running time:} Tests take of the order of several minutes;
Otherwise highly problem dependent (from minutes to several days).

\end{small}


\newcommand{\purpose}[1]{\textcolor{red}{\it #1: }}
\newcommand{\todo}[1]{\textcolor{blue}{\it #1.}}
\section{Introduction}
\label{sec:introduction}
The open source ALPS (Algorithms and Libraries for Physics Simulations) project \cite{ALPS10,ALPS13,ALPS2.0} provides a collection of physics libraries and applications, especially suited for the simulation of lattice models and strongly correlated systems. The ALPS libraries provide well-documented, reusable components for developing condensed matter physics simulation code, while ALPS applications' goal is to make commonly used and proven computational algorithms available to a non-expert community.

Computer codes based on the ALPS libraries have provided physical insights in many subfields of condensed matter. Highlights include nonequilibrium dynamics \cite{Eckstein09,Werner09}, continuous-time quantum Monte Carlo \cite{Werner06,Gull08}, LDA+DMFT materials simulations \cite{Ferber12}, simulations of quantum \cite{Cremades12} and classical \cite{Bergman07} spin, correlated boson \cite{Pollet10} and fermion \cite{LeBlanc15} models, as well as cuprate superconductivity \cite{Gull13}.

In this paper we present an updated and refactored version of the main ALPS Libraries (here and thereafter referred to as ``ALPS core libraries'' or, as a whole, ``the library''). The focus of the update is on providing a set of lightweight, thoroughly tested and documented libraries to code developers that implement generic algorithms and utilities for rapid development of efficient computational physics applications. Substantial portions of the code are significantly rewritten compared to the previous version of the ALPS Libraries, and special emphasis is placed on a short development cycle and up-to-date user documentation.

A snapshot of the code at the time of submission of this paper is available at the CPC program library. Later versions will be available from the main site, \url{http://alpscore.org}, together with in-depth user documentation, tutorials, and examples. Development versions of the library are available from the public \Git repository at \mbox{\url{https://github.com/ALPSCore/ALPSCore}}.  Several application codes are built on the core libraries and are distributed independently.

The remainder of this paper is organized as follows: In section~\ref{sec:motphil} we introduce and motivate the library; in section~\ref{sec:components},
we provide an overview of the library  components; in
section~\ref{sec:prereqs}, we list the software packages prerequisite
for compiling and installing the library, as well as outline the
installation procedure; in section~\ref{sec:devtest}, we describe the
development and testing cycle of the library; in
section~\ref{sec:cite} we specify the ALPS citation policy and license, and in sections~\ref{sec:summary} and~\ref{sec:ackn} we
summarize the paper and acknowledge our contributors. Additionally, \ref{sec:installation-detail} provides installation examples and \ref{sec:example-ising} an example of code using
the library to program a two-dimensional Ising model simulation with single spin flip updates.

\section{Motivation, target audience, philosophy}\label{sec:motphil}
Our library is intended to be used by two communities: directly by computational physicists who develop their own computer codes, and indirectly by users who use applications based on the library. There are a number of challenges 
that such a scientific developer faces:

\paragraph{Framework/boilerplate code} A typical computational
physics program (for example, a Monte Carlo simulation code) contains
parts that have little to do with the scientific problem that the
simulation strives to get insight into. Those parts form the necessary
framework for running the simulation, and are responsible for various
auxiliary tasks such as reading the parameters of the simulation or
saving intermediate and final results. Auxiliary `boilerplate' code is
often imported with little or no changes from application
to application. As writing the auxiliary
code is secondary to solving the scientific problem
at hand, it is therefore given less attention. However, the auxiliary code
is not necessarily trivial and redeveloping it from scratch results in
unnecessary duplication of effort and reduced code quality.

\paragraph{Code reuse problem} For the same reasons, physics code often
lacks modularization: parts of the code are not properly
separated by well-defined interfaces, and the scientific
code is buried deep inside the boilerplate/framework. The
resulting lack of interoperability hinders the exchange of
code with other research groups and results in further
unnecessary duplication of effort. 

\paragraph{Process management} As codes grow, they become progressively
more difficult to manage and document. The lack of modularity and
documentation makes it hard for researchers outside the
primary developers group to contribute to the project. A possibility
of unintentional disruption of code functionality by introducing
changes effectively discourages implementing new features.

\paragraph{High performance computing requirements} Although
programming in a high-level language (such as \Python) may help in
solving the framework/boilerplate code problem, the use of a
high-level language is often problematic. Computational physics
algorithms are complex, and the computational kernels implementing
them are resource-demanding and in most of the cases must make use of
multi-node and multi-core capabilities of contemporary HPC (High Performance Computing)
hardware. The high-performance requirements necessitate implementation
of the kernels in a lower-level language, such as \CXX, and utilizing
\MPI and \OpenMP parallel programming techniques. The task of including
high-performance parallelized kernels into high-level language
framework, although technically feasible, is far from being trivial.

\vspace{1\baselineskip}
The purpose of the ALPS core libraries is to reduce the user's
time and effort to develop and test complex scientific applications.
This project addresses the above-mentioned challenges by providing
reusable and widely-used, high-performance, well-documented software
building blocks with open architecture. To facilitate reuse, the
library utilizes high level of abstraction without compromising
performance. To encourage reuse and contributions, the library strives
to maintain current user-level and developer-level documentation, easy
installation procedure, as well as a short development cycle with
rapid bugfixes and frequent releases. The manageability
requirements are addressed via documented code, modular structure, and
extensive testing. The project is open to contributions in the form of
bug reports, feature requests, and code.

\section{Some of the key library components}
\label{sec:components}

In this section, we provide a brief overview of the purpose and functionality of some of
the key components comprising the ALPS core libraries. 
Additional information is
available online and as part of the \Doxygen code documentation.
It should be noted that the components maintain minimal interdependence, and 
using a component in a program does not bring in other components, unless they are 
required dependencies, in which case they will be used automatically.

\paragraph{Building a parallel Monte Carlo simulation} A generic Monte Carlo simulation
can be easily assembled from the classes provided by \emph{Monte
  Carlo Scheduler} component. The programmer needs to define only the
problem-specific methods that are called at each Monte Carlo step,
such as methods to update the configuration of the Monte Carlo chain and to
collect the measured data. The simulation is parallelized
implicitly, using one \MPI process per
chain. \ref{sec:example-ising} contains an example of a Monte Carlo
simulation using these Scheduler classes.

\paragraph{Computing observable averages, errors, correlation
  times, and cross-correlation} To collect Monte Carlo simulation data
and to estimate and propagate Monte Carlo errors of potentially correlated data, the
\emph{Accumulators} library component is to be used, especially in
cases where correlations in time or cross-correlations between
different variables exist. The Accumulator objects support arithmetic
operations and elementary mathematical functions, e.g. division of one
observable by another or the multiplication of two observables. The non-linear error 
propagation through these operations is also supported.

Several Accumulator objects provide different
trade-offs between memory consumption, overhead time, and functionality.
In order of decreasing memory and measurement requirements they are:
\begin{enumerate}
\item \verb|FullBinningAccumulator|: provides access to data
  auto\-correlation length, auto\-correlation-corrected error bars, and
  proper error propagation in non-linear functions using Jackknife~\cite{jackknife} resampling technique.
\item \verb|LogBinningAccumulator|: provides access to data
  auto\-correlation estimates and auto\-correlation-corrected error bars.
\item \verb|NoBinningAccumulator|: estimates the error bars of the
  data without regard to auto\-correlation.
\item \verb|MeanAccumulator|: maintains only the mean value of the
  data (no error bars or auto\-correlation length).
\end{enumerate}
A typical use of accumulators in a Monte Carlo simulation is
demonstrated in~\mbox{\ref{sec:example-ising}}.

\paragraph{Storing, restoring, and checkpointing simulation results} To store the results of
a simulation in a cross-platform format for subsequent analysis, one can
use the \emph{Archive} component. The component provides convenient
interface to saving and loading of common \CXX data structures
(primitive types, complex numbers, STL vectors and maps), as well as
of objects of user-defined classes \HDF~\cite{hdf5}, which is the universally supported and
machine independent 
data format. Use of the Archive
component for application checkpointing is illustrated
in~\mbox{\ref{sec:example-ising}}.

\paragraph{Reading command-line arguments and parameter files} Input
parameters to a simulation can be passed via a combination of a
parameter file and command line arguments. The \emph{Parameters}
library component is responsible for parsing the files and the command
line, and providing access to the data in the form of an
associative array (akin to \CXX \verb|map| or \Python dictionary). The
parameter files use the standard ``\verb|*.ini|'' format, a plain text
format with a line-based syntax containing \verb|key = value| pairs,
optionally divided into sections. The use of the Parameters component
is illustrated in~\ref{sec:example-ising}.

\paragraph{Working with Green's functions} The \emph{Green's
  Functions} component provides a type-safe interface to manipulate
objects representing bosonic or fermionic many-body Green's functions,
self-energies, susceptibilities, polarization functions, and similar
objects. From a programmer's perspective, these objects are
multidimensional arrays of floating-point or complex numbers, defined
over a set of meshes and addressable by a tuple of indices, each
belonging to a grid.  Currently, Matsubara (imaginary frequency),
imaginary time, power, momentum space, real space, and arbitrary index
meshes are supported.

These many-body objects often need to be supplemented with analytic tail information
encapsulating the high frequency / small time moments of the Green's
functions, so that high precision Fourier transforms, density
evaluations, or energy evaluations can be performed. The Green's
function library component provides this functionality in addition to
saving to and loading from binary \HDF files. An example illustrating
the use of the Green's function classes is given
in~\ref{sec:example-gf}.

\section{Prerequisites and Installation}
\label{sec:prereqs}
To build the ALPS core libraries, any recent \CXX compiler can be
used; the libraries are tested with \GCC \cite{gcc} 4.2 and above,
Intel \cite{icc} \CXX 10.0 and above, and Clang \cite{clang} 3.2 and
above. The library follows the \CXX{}03
standard \cite{cpp03} to facilitate the portability to a
wide range of programming environments, including HPC clusters with older compilers. 
The library depends on the following external packages:
\begin{itemize}
\item The \CMake build system \cite{cmake} of version 2.8.12 and above.
\item The \Boost \CXX libraries \cite{boost} of version 1.54.0 and above. Compiled libraries are only needed from the program option, serialization, and file system libraries. 
\item The \HDF library \cite{hdf5} version 1.8 and above.
\end{itemize}
To make use of (optional) parallel capabilities, an \MPI implementation
supporting standard 2.1~\cite{mpi-2.1} and above is required. Generating the developer's documentation requires \Doxygen
\cite{doxygen}  along with its dependencies.

The installation of the ALPS core libraries follows the standard procedure for any
\CMake-based package. The first step is to download the ALPS core libraries source
code; the recommended way is to download the latest ALPS core libraries release
from \url{https://github.com/ALPSCore/ALPSCore/releases}. Assuming
that all above-mentioned prerequisite
software is installed, the installation consists of unpacking the
release archive and running \CMake from a temporary build directory, as
outlined in the shell session example below (the \verb|$| sign designates a shell prompt):
\begin{bashlisting}[emph={tar,mkdir,cmake,make},emphstyle={\color{darkgreen}}]
$ tar -xzf ALPSCore-0.5.4.ta#()r.gz
$ mkdir build
$ cd build
$ export ALPSCore_DIR=$HOME/software/ALPSCore
$ cmake -DCMAKE_INSTALL_PREFIX=$ALPSCore_DIR \
        -DCMAKE_BUILD_TYPE=Release \
        ../ALPSCore-0.5.4 
$ make
$ make t#()est
$ make install
\end{bashlisting} 
The command at line~1 unpacks the release archive (version~0.5.4 in
this example); at line~4 the destination install directory of the
ALPS core libraries is set (\verb|$HOME/software/ALPSCore| in this
example). 

The ALPS core libraries come with an extensive set of tests;
it is strongly recommended to run the tests (via
\verb|make test|) to verify the
correctness of the build, as it is done at line~9 in the example
above.

The installation procedure is outlined in more details
in~\ref{sec:installation-detail}; also, the file \verb|common/build/build.jenkins.sh| in the library release
source tree contains a build and installation script that can be further
consulted for various build options. 

On Mac~OS~X operating system, the ALPS core libraries package can be downloaded
and installed from the \code{\mbox{MacPorts}}~\cite{macports} or \code{\mbox{HomeBrew}}~\cite{homebrew}
repositories, using commands \verb|port install alpscore| or
\verb|brew tap homebrew/science && brew install alpscore|, respectively.

\section{Development and Test Cycle}
\label{sec:devtest}
The ALPS core libraries development emphasizes a short development cycle
and frequent releases of operational code, along with user-level and
programmer-level documentation. The techniques we employ to achieve these
goals are discussed below.
\subsection{Distributed version control and  collaborative
  development}
In accordance with modern software practices, all changes introduced
in the course of development of ALPS core libraries are tracked using
a distributed version control system, \Git~\cite{git}. The
version control system preserves the history of changes in every
source file in its database (the \Git repository), associating
each change (a ``commit'') with its originator, and allows the
author to comment on each set of changes. The version control
system allows one to create independent lines of code development
(branches), with subsequent optional merging of those lines.

The master version of the repository with ALPS core libraries, which
is publicly available for downloading, is hosted on the
GitHub~\cite{github} hosting service. The distributed nature of
\Git allows any developer to work independently on his or her
own copy (``clone'') of the repository, and then request to merge his
or her clone into this publicly-visible repository (submit a ``pull
request''); after reviewing the code changes that the developer
made, the request is granted by one of the members of the core
development team.

In addition to the repository, GitHub provides an issue tracker. Any
registered GitHub user is able to submit a question, a bug report or a
feature request via the tracking system. The developer team is
notified that a new issue is open and then triages the issue, marking
it according to its severity (for example, ``bug'', ``enhancement'',
``question''); assigns a developer to be responsible for it; and
associates a code development ``milestone'' with it.  The issue can be
further commented on by other users and developers, creating a thread
of conversation. Once the issue is resolved or otherwise addressed, it
is marked as ``closed''.  

GitHub also provides hosting space for Wiki pages, which the project
uses for user-level documentation and tutorials. The Wiki
documentation contains direct links to repository code, and thus
simplifies keeping the code samples up to date. The ALPS core libraries
project encourages collaboration and feedback from the user community by
opening issues, submitting pull requests, and contributing to the Wiki
documentation pages.


\subsection{Test-driven development and continuous integration} 

Test-driven development (TDD) is a software development process that
emphasizes creating a test for a feature prior to implementing
it. Among the advantages of TDD are increased test coverage (because
every new feature has a corresponding test), higher confidence in test
validity (because the test is ensured to fail on unimplemented
feature) and improved code quality~\cite{beck03:TDD}. We follow the
TDD process in the development of the ALPS core libraries, whenever
practical.

Typically, each test verifies the behavior of a single function or
object method in isolation from other methods of the same class; this
approach is known as ``unit testing''. However, many components of
ALPS core libraries depend on functionality provided by other 
components (for example, Monte Carlo scheduler uses
Accumulators and Parameters). In this way, a unit test of an object
method provided by a component tests the coordinated working of the
component's dependencies, and can be considered a form of integration
test. It should also be noted that only publicly accessible (rather
than private) methods are invoked in testing. In our development
process, a unit tests is also a form of documentation, demonstrating both
the correct way to call a method and the interface specification.

In order to avoid maintenance problems associated with merging several
branches of development and possible breaking of code functionality,
the ALPS core libraries development adopted the Continuous
Integration~\cite{beck04:CI} practice. Continuous Integration
prescribes frequent merging of the code into the main repository and
running tests (in our case, the unit/integration test suite) to make
sure that the code functions as expected. To build the code and run
the test suite we utilize the Jenkins Continuous Integration
tool~\cite{jenkins}. In order to conduct unit and integration testing
of the ALPS core libraries on wider range of operating environments,
we make use of a cloud-based service (provided by
CloudBees~\cite{cloudbees}). When a developer merges his or her series
of commits into the central repository, an automated upload to the
cloud-based service occurs, followed by building of the library and
running the test suites in different environments, under different
operating systems (Linux and Mac~OS~X) and using various versions of
compilers and libraries. The test results are immediately available
for developers to analyze and correct any errors. The same build
process also generates developer-level documentation from documenting
comments embedded in the code.

\begin{figure}[h!]
\centering
\includegraphics[width=0.8\textwidth]{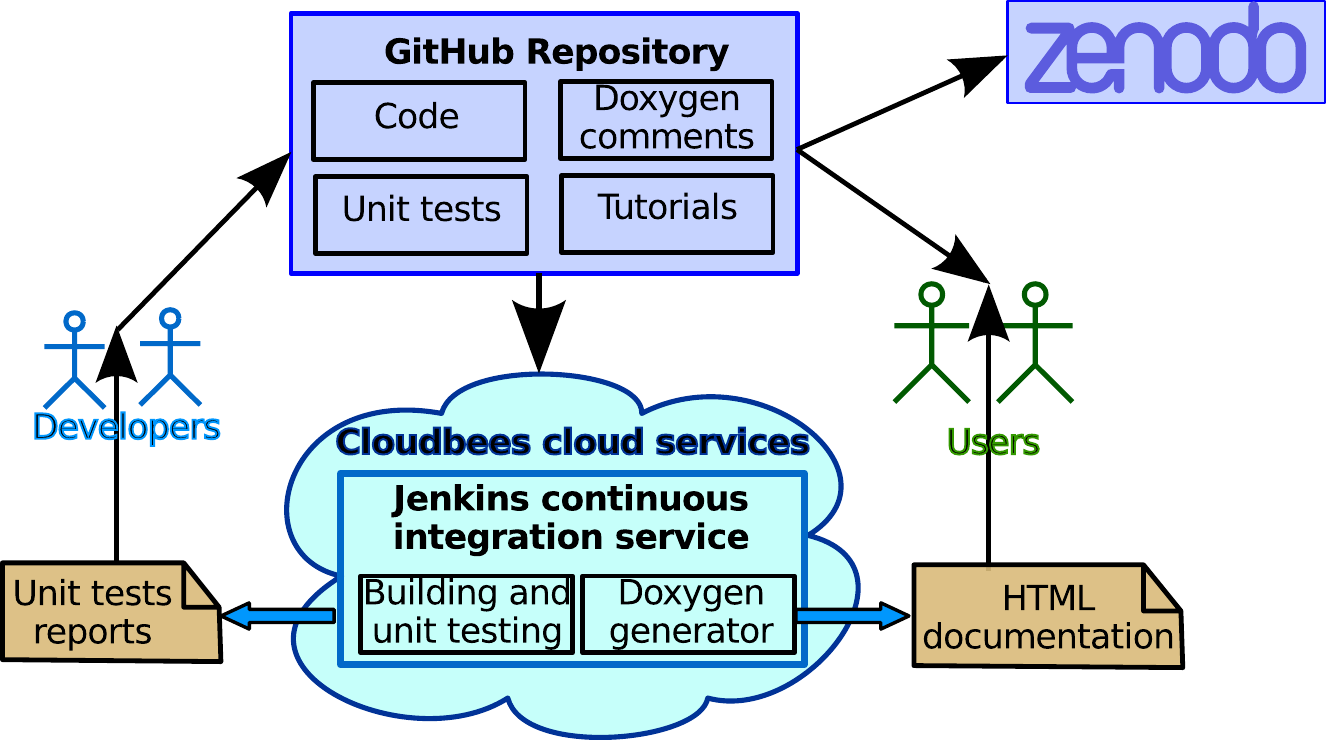}
\caption{Development cycle of ALPS core libraries. Developer's commit to
  public GitHub repository causes automatic rebuild on CloudBees
  service via Jenkins. Releases are automatically published on
  Zenodo.\label{fig:dev-workflow}}
\end{figure}

Once all outstanding issues are resolved, new features are implemented
or existing ones are improved, and the test suites are passed in all
supported environments, a developer requests GitHub to generate a
release.  By virtue of the Continuous Integration process, the code in
repository at that stage is ensured to be free of known issues;
therefore, the release is essentially a compressed archive of the
snapshot of the repository, ready to be installed and deployed. When
the state of the repository is marked as a release, a request is
automatically sent to the Zenodo~\cite{zenodo} service to download and
archive the release in Zenodo's searchable database; Zenodo also
generates a citable DOI code associated with the release. The complete
development and test cycle of the ALPS core libraries project,
discussed above, is illustrated by Figure~\ref{fig:dev-workflow}.

\section{License and citation policy}
\label{sec:cite}
The GitHub version of ALPS core libraries is licensed under the GNU General Public
License version 2 (GPL~v.~2)~\cite{gplv2} or later; for compatibility
reasons with the journal the version released here is licensed under
GPL~v.~3. The older ALPS license under which previous versions of the
code were licensed \cite{ALPS2.0} has been retired. We kindly request
that the present paper be cited, along with any relevant original
physics or algorithmic paper, in any published work utilizing an application code
that uses this library.

\section{Summary}
\label{sec:summary}
We have presented an updated and repackaged version of the core ALPS libraries, a lightweight \CXX library, designed to
facilitate rapid development of computational physics applications, 
and have described its main
features. The collaborative, test-driven development process utilizing
the continuous integration approach has been also described.

\section{Acknowledgments}
\label{sec:ackn}
Work on the ALPS library project is supported by the Simons collaboration on the many-electron problem. Aspects of the library are supported by NSF DMR-1606348 and DOE ER 46932. 
\section{References}

\bibliographystyle{elsarticle-num}
\bibliography{references}

\appendix

\section{Detailed installation procedure}
\label{sec:installation-detail}
In the following discussion we assume that all
prerequisite software (section~\ref{sec:prereqs}) is installed, and the ALPS core libraries
release (here, release~0.5.4) is downloaded into the current directory
as \verb|ALPSCore-0.5.4.tar.gz|. Also, we assume that the libraries are
to be installed in \verb|ALPSCore| subdirectory of the current user's
home directory. The commands are given assuming \verb|bash| as a user
shell. 

The first step is to unpack the release archive and set the desired
install directory:
\begin{bashlisting}[emph={tar,mkdir,cmake,make},emphstyle={\color{darkgreen}}]
$ tar -xzf ALPSCore-0.5.4.t#()ar.gz
$ export ALPSCore_DIR=$HOME/software/ALPSCore
\end{bashlisting} 

The next step is to perform the build of the library
(note that the build should not be performed in the source directory):
\begin{bashlisting}[emph={tar,mkdir,cmake,make},emphstyle={\color{darkgreen}},firstnumber=last]
$ mkdir build
$ cd build
$ cmake ../ALPSCore-0.5.4 -DCMAKE_INSTALL_PREFIX=$ALPSCore_DIR \
                          -DCMAKE_BUILD_TYPE=Release
\end{bashlisting}

The \verb|cmake| command at lines~5 and~6 accepts additional arguments
in the format \code{-D\textit{variable}=\textit{value}}. A number of
relevant \CMake variables is listed in Table~\ref{tab:cmake-args}. The
installation process is also affected by environment variables, some
of which are listed in Table~\ref{tab:cmake-env}; 
the \CMake variables take precedence over the environment variables.
The build and
installation script \verb|common/build/build.jenkins.sh| in the
ALPS core libraries release source tree provides an example of using some of the
build options.

\begin{table}
  \centering
  \begin{tabularx}{\textwidth}{lc>{\raggedright\arraybackslash}X}
    \textbf{Variable} & \textbf{Default value} & {\hfil\textbf{Comment}\hfil}\\
    \toprule
    \code{CMAKE\_CXX\_COMPILER} & (system default) & {Path to \CXX compiler executable.*} \\\midrule
    \code{CMAKE\_INSTALL\_PREFIX} & \code{/usr/local} & library target install directory. \\\midrule
    \code{CMAKE\_BUILD\_TYPE} &  &  {Specifies build type;
                                  set to \code{Release} to maximize performance.} \\\midrule
    \code{BOOST\_ROOT} &   & {\Boost install directory.
                            Set if \CMake fails to find \Boost.} \\\midrule
    \code{Boost\_NO\_SYSTEM\_PATHS} & \code{false} & {Set to \code{true} to disable search in default system directories,
                                           if the wrong version of \Boost is found.} \\\midrule
    \code{Boost\_NO\_BOOST\_CMAKE} & \code{false} & {Set to \code{true} to disable search for \Boost \CMake file,
                                          if the wrong version of \Boost is found.} \\\midrule
    \code{Documentation} & \code{ON} & Build developer's documentation. \\\midrule
    \code{ENABLE\_MPI} & \code{ON} & Enable \MPI build (set to \code{OFF} to disable). \\\midrule
    \code{Testing} & \code{ON} & Build unit tests (recommended). \\\midrule
    \code{ALPS\_BUILD\_TYPE} & \code{dynamic} & {Can be \code{dynamic} or
      \code{static}: build libraries as dynamic (``shared'') or static
      libraries, respectively.* \emph{(Since ALPSCore 0.5.5)}} \\\midrule
    \multicolumn{3}{c}{\emph{For ALPSCore version 0.5.4 and older:}}\\\midrule
    \code{ALPS\_BUILD\_SHARED} & \code{ON} & {Build shared ALPS core libraries.*
                                    Mutually exclusive with \code{ALPS\_BUILD\_STATIC=ON}.} \\\midrule
    \code{ALPS\_BUILD\_STATIC} & \code{OFF} & {Build static ALPS core libraries.*
                                    Mutually exclusive with \code{ALPS\_BUILD\_SHARED=ON}.} \\
    \bottomrule
    \multicolumn{3}{l}{{}*\footnotesize Note: For the change of this variable
        to take effect, remove your build directory and redo the build.}
  \end{tabularx}
  \caption{\CMake arguments relevant to building of ALPS core libraries.}
  \label{tab:cmake-args}
\end{table}

\begin{table}
  \centering
  \begin{tabularx}{1\textwidth}{l@{\hspace{5em}}X}
    \textbf{Variable} & {\hfil\textbf{Comment}\hfil}\\
    \toprule
    \code{CXX} &  Path to \CXX compiler executable.* \\\midrule
    \code{BOOST\_ROOT} &  \Boost install directory.
                        Set if \CMake fails to find \Boost. \\\midrule
    \code{HDF5\_ROOT}  &  \HDF install directory.
                        Set if \CMake fails to find \HDF.\\

    \bottomrule
    \multicolumn{2}{l}{{}*\footnotesize Note: For the change of this variable
        to take effect, remove your build directory and redo the build.}
  \end{tabularx}
  \caption{Environment variables arguments relevant to building of ALPS core libraries.}
  \label{tab:cmake-env}
\end{table}

\section{Example: 2D Ising model}
\label{sec:example-ising}

\subsection{Problem statement}

This program implements a single-spin update Metropolis Monte Carlo
algorithm for computing properties of a well-known two-dimensional
Ising model. The system under consideration is a square lattice of
spins $\{s_{i}=\pm 1\}$; the periodic boundary conditions are
imposed, with a unit cell of size $L\times L$, containing $N=L^{2}$
spins. The energy per spin is given by
\begin{equation}
  \label{eq:ising-energy}
  E = - \frac{1}{N}\sum_{(i,j)} s_{i}s_{j}
\end{equation}
where $(i,j)$ runs over all nearest-neighbor pairs in the unit
cell. The magnetization per spin is given by
\begin{equation}
  \label{eq:ising-magnetization}
  M = \frac{1}{N} \sum_{i} s_{i}
\end{equation}
where $i$ runs over all $N=L^{2}$ spins in the unit cell. 

The properties of interest are thermal averages of energy $\langle E
\rangle$, magnetization $\langle M \rangle$, absolute value of
magnetization $\langle |M| \rangle$ and Binder cumulant:
\begin{equation}
  \label{eq:ising-binder-cumulant}
  U = 1 - \frac{\langle M^{4} \rangle}{3 \langle M^{2} \rangle^{2}}
\end{equation}
computed at temperature $T$ (inverse temperature $\beta=1/T$)
expressed in energy units. 


\subsection{Build process}
The build of the example program is controlled by the
\verb|CMakeLists.txt| file given by the
Listing~\ref{cmake-example}. 

\begin{lstlisting}[caption=\textbf{CMakeLists.txt} file, label=cmake-example,numbers=right,language=sh,emph={cmake_minimum_required,project,add_executable,find_package,target_link_libraries},emphstyle={\color{darkgreen}}]
cmake_minimum_required(VERSION 2.8.12)
project(ising2_mc CXX)

add_executable(${PROJECT_NAME} main.cpp ising.cpp)
add_executable(${PROJECT_NAME}_mpi main_mpi.cpp ising.cpp)

# Request the ALPSCore package (with all components)
# The ALPSCore package must be installed in some
# standard place (like /usr/local),
# or somewhere in your PATH. 
# Otherwise, point ALPSCore_DIR environment variable
# to the ALPScore installation tree.
find_package(ALPSCore REQUIRED)

# Use ALPSCore_LIBRARIES variable to link to ALPSCore 
target_link_libraries(${PROJECT_NAME} ${ALPSCore_LIBRARIES})
target_link_libraries(${PROJECT_NAME}_mpi ${ALPSCore_LIBRARIES})
\end{lstlisting}

Lines~4 and~5 specify the source files
that constitute the sequential and parallel versions of the program,
respectively. Line~13 requests the ALPS core libraries (ALPSCore) package that must be
installed as described in section~\ref{sec:installation-detail}; the
installation location must be in one of standard system directories
such as \verb|/usr/local|, in one of the directories listed in the
\verb|PATH| environment variable, or in a directory pointed to by
\verb|ALPSCore_DIR| environment variable. Lines~16 and~17 specify that
the sequential and parallel versions of the program use libraries
provided by the ALPSCore package.

Assuming that the ALPS core libraries are installed as described above,
the example code is located in \emph{/path/to/source} directory, and
the project is to be built in \emph{/path/to/build} directory, the
following commands will build the project (the \verb|$| sign
designates a shell prompt):
\begin{bashlisting}[emph={tar,mkdir,cmake,make},emphstyle={\color{darkgreen}}]
$ cd #(\textit{/path/to/build})
$ cmake #(\textit{/path/to/source})
$ make
\end{bashlisting} 
Two executable files, \verb|ising2_mc| for the sequential version, and
\verb|ising2_mc_mpi| for the parallel version, will be generated in
the current directory.

\subsection{Monte Carlo simulation class}

The simulation class encapsulates the details of building a Markov
chain, sampling observables (in this case, the energy, magnetization, \emph{etc.}),
and keeping track of the progress of the calculation. Our simulation
class \verb|ising_sim| is declared in
Listing~\ref{ising-hpp-example}, line~11. 

\begin{lstlisting}[caption={\textbf{ising.hpp} header file: Declares interface of
the simulation class}, label=ising-hpp-example,numbers=right]
#pragma once

#include <alps/mc/mcbase.hpp>
#include "storage_type.hpp"


// Simulation class for 2D Ising model (square lattice).
// Extends alps::mcbase, the base class of all Monte Carlo simulations.
// Defines its state, calculation functions (update/measure) and
// serialization functions (save/load)
class ising_sim : public alps::mcbase {
    // The internal state of our simulation
  private:
    int length; // the same in both dimensions
    int sweeps;
    int thermalization_sweeps;
    int total_sweeps;
    double beta;
    storage_type spins;
    double current_energy;
    double current_magnetization;



  public:
    ising_sim(parameters_type const & parms, std::size_t seed_offset = 0);

    static void define_parameters(parameters_type & parameters);

    virtual void update();
    virtual void measure();
    virtual double fraction_completed() const;

    using alps::mcbase::save;
    using alps::mcbase::load;
    virtual void save(alps::hdf5::archive & ar) const;
    virtual void load(alps::hdf5::archive & ar);
};

\end{lstlisting}

The simulation class is required to inherit from \verb|alps::mcbase|,
and it must define the following virtual methods with the
corresponding semantics:
\begin{itemize}
\item \verb|void update()| (line~30): Attempt a step in the phase
  space, accept or reject it;
\item \verb|void measure()| (line~31): Compute and accumulate
  observables;
\item \verb|double fraction_completed() const|  (line~32): report the
  fraction of the computation that is completed.
\end{itemize}
Also, a static method \verb|void define_parameters()| (line~28) should
be provided that declares (``defines'') the input parameters pertinent
to the simulation (see section~\ref{sec:components}).

Listing~\ref{ising-hpp-example} also declares a number of
implementation-specific private data members (lines~14--23) holding
the internal state of the Markov chain. The \code{include} statement
in line~4 refers to the header file containing implementation
details: storage of spins in a two-dimensional array; for the sake of
completeness, the implementations of the corresponding class is
briefly discussed in subsection~\ref{subsec:impl-details}.

The implementation of the simulation class is provided by
\code{ising.cpp} file.\footnote{The code presented here and in the
following discussion is written with the focus on simplicity and clarity rather than efficiency.} 
Lines~5--21 in the implementation file
(Listing~\ref{ising-cpp-example-defpar}) define the input parameters
that are relevant to the simulation: size of the unit cell, number of
thermalization Monte Carlo steps, total number of steps, and the
simulation temperature. Note that each parameter name has an
associated type, and parameters ``sweeps'' and ``thermalization'' are
given default values (0 and 10000, respectively); the other two
parameters (``length'' and ``temperature'') do not have default values
and are required to be provided by the user. In addition, a number of
standard parameters used in every Monte Carlo simulation is defined by
the base class; the corresponding static method is called at
line~12. Line~15 defines a number of parameters common for many
simulations.

\begin{lstlisting}[caption={\textbf{ising.cpp} file: Implementation of the
  simulation class: Parameter definition},
label=ising-cpp-example-defpar,numbers=right,lastline=22]
#include "ising.hpp"
#include <alps/params/convenience_params.hpp>

// Defines the parameters for the ising simulation
void ising_sim::define_parameters(parameters_type & parameters) {
    // If the parameters are restored, they are already defined
    if (parameters.is_restored()) {
        return;
    }
    
    // Adds the parameters of the base class
    alps::mcbase::define_parameters(parameters);
    // Adds the convenience parameters (for save/load)
    // followed by the ising specific parameters
    alps::define_convenience_parameters(parameters)
        .description("2D ising simulation")
        .define<int>("length", "size of the periodic box")
        .define<int>("sweeps", 0, "maximum number of sweeps (0 means indefinite)")
        .define<int>("thermalization", 10000, "number of sweeps for thermalization")
        .define<double>("temperature", "temperature of the system");
}
\end{lstlisting}

The object constructor (Listing~\ref{ising-cpp-example-ctor})
initializes the internal state of the simulation in
lines~27--56. Additionally, lines~59--66 allocate named accumulators
for the observables to be collected during the simulation.

\begin{lstlisting}[caption={\textbf{ising.cpp} file: Implementation
  of the simulation class: Constructor}, label=ising-cpp-example-ctor,numbers=right,firstline=23,firstnumber=23,lastline=67]
// Creates a new simulation.
// We always need the parameters and the seed as we need to pass it to
// the alps::mcbase constructor. We also initialize our internal state,
// mainly using values from the parameters.
ising_sim::ising_sim(parameters_type const & parms, std::size_t seed_offset)
    : alps::mcbase(parms, seed_offset)
    , length(parameters["length"])
    , sweeps(0)
    , thermalization_sweeps(int(parameters["thermalization"]))
    , total_sweeps(parameters["sweeps"])
    , beta(1. / parameters["temperature"].as<double>())
    , spins(length,length)
    , current_energy(0)
    , current_magnetization(0)
{

    // Initializes the spins
    for(int i=0; i<length; ++i) {
        for (int j=0; j<length; ++j) {
            spins(i,j) = (random() < 0.5 ? 1 : -1);
        }
    }

    // Calculates initial magnetization and energy
    for (int i=0; i<length; ++i) {
        for (int j=0; j<length; ++j) {
            current_magnetization += spins(i,j);
            int i_next=(i+1)%length;
            int j_next=(j+1)%length;
            current_energy += -(spins(i,j)*spins(i,j_next)+
                                spins(i,j)*spins(i_next,j));
            
        }
    }
    
    // Adds the measurements
    measurements
        << alps::accumulators::FullBinningAccumulator<double>("Energy")
        << alps::accumulators::FullBinningAccumulator<double>("Magnetization")
        << alps::accumulators::FullBinningAccumulator<double>("AbsMagnetization")
        << alps::accumulators::FullBinningAccumulator<double>("Magnetization^2")
        << alps::accumulators::FullBinningAccumulator<double>("Magnetization^4")
        ;
}
\end{lstlisting}

Implementation of the Monte Carlo step and the acceptance/rejection
logic by the method \verb|update()| (Listing~\ref{ising-cpp-example-update},
lines~70--97) is straightforward.

\begin{lstlisting}[caption={\textbf{ising.cpp} file: Implementation
  of the simulation class: Update method}, label=ising-cpp-example-update,numbers=right,firstline=68,firstnumber=68,lastline=98]
// Performs the calculation at each MC step;
// decides if the step is accepted.
void ising_sim::update() {
    using std::exp;
    typedef unsigned int uint;
    // Choose a spin to flip:
    uint i = uint(length * random());
    uint j = uint(length * random());
    // Find neighbors indices, with wrap over box boundaries:
    uint i1 = (i+1) % length;            // right
    uint i2 = (i-1+length) % length;     // left
    uint j1 = (j+1) % length;            // up
    uint j2 = (j-1+length) % length;     // down
    // Energy difference:
    double delta=2.*spins(i,j)*
                    (spins(i1,j)+  // right
                     spins(i2,j)+  // left
                     spins(i,j1)+  // up
                     spins(i,j2)); // down
    
    // Step acceptance:
    if (delta<=0. || random() < exp(-beta*delta)) {
        // update energy:
        current_energy += delta;
        // update magnetization:
        current_magnetization -= 2*spins(i,j);
        // flip the spin
        spins(i,j) = -spins(i,j);
    }        
}
\end{lstlisting}

The \verb|measure()| method (Listing~\ref{ising-cpp-example-measure},
lines~100--113) adds the observables of interest to the corresponding
named accumulators (lines~108--112).

\begin{lstlisting}[caption={\textbf{ising.cpp} file: Implementation
  of the simulation class: Measurement method}, label=ising-cpp-example-measure,numbers=right,firstline=99,firstnumber=99,lastline=114]
// Collects the measurements at each MC step.
void ising_sim::measure() {
    ++sweeps;
    if (sweeps<thermalization_sweeps) return;
    
    const double n=length*length; // number of sites
    double tmag = current_magnetization / n; // magnetization

    // Accumulate the data (per site)
    measurements["Energy"] << (current_energy / n);
    measurements["Magnetization"] << tmag;
    measurements["AbsMagnetization"] << fabs(tmag);
    measurements["Magnetization^2"] << tmag*tmag;
    measurements["Magnetization^4"] << tmag*tmag*tmag*tmag;
}
\end{lstlisting}

The \verb|fraction_completed()| method
(Listing~\ref{ising-cpp-example-frac}, lines~116--122) returns a
\verb|double| value. This value indicates whether the simulation ran the intended
number of steps: the simulation stops when the value reaches
$1.0$. The termination criteria are discussed in more details
in~\ref{subsec:scheduling}.

\begin{lstlisting}[caption={\textbf{ising.cpp} file: Implementation
  of the simulation class: Report completed fraction},
label=ising-cpp-example-frac, numbers=right,firstline=115,firstnumber=115,lastline=123]
// Returns a number between 0.0 and 1.0 with the completion percentage
double ising_sim::fraction_completed() const {
    double f=0;
    if (total_sweeps>0 && sweeps >= thermalization_sweeps) {
        f=(sweeps-thermalization_sweeps)/double(total_sweeps);
    }
    return f;
}
\end{lstlisting}

The implementation file \code{ising.cpp} also contains definitions of methods that are
necessary for checkpointing the simulation, as discussed
in~\ref{subsec:checkpointing}.

\subsection{Main program}

We first describe the flow of the sequential implementation,
postponing the discussion of the parallelization till
subsection~\ref{subsec:scheduling}. The sequential main program
(\code{main.cpp}) is shown in Listings~\ref{main-cpp-example-start}
and~\ref{main-cpp-example-collect}; a few essential points are
discussed in this and the following subsections. 

Line~19 of Listing~\ref{main-cpp-example-start} constructs a
parameter object from the command line arguments: an optional name
of a parameter file followed by an optional set of parameters in the
form of ``\verb|--key=value|''. The parameter values that are set in the command line
override the values set in the parameter file.
The parameter names and types must be
declared before use, which is done at line~20 by a call to a static
method \verb|define_parameters()| of the simulation
class. Lines~22--25 check if the user requested help or missed a
required argument. The simulation is created at line~28, with the
parameters passed to the constructor. Line~41 starts the simulation.

\begin{lstlisting}[caption={\textbf{main.cpp} file: Main program, sequential version: Starting the simulation},
 label=main-cpp-example-start, numbers=right, lastline=42]
#include "ising.hpp"
#include <iostream>
#include <alps/accumulators.hpp>
#include <alps/mc/api.hpp>
#include <alps/mc/mcbase.hpp>
#include <alps/mc/stop_callback.hpp>

int main(int argc, char* argv[])
{
    // Define the type for the simulation
    typedef ising_sim my_sim_type;

    try {
    
        // Creates the parameters for the simulation
        // If an hdf5 file is supplied, reads the parameters there
        std::cout << "Initializing parameters..." << std::endl;

        alps::params parameters(argc, (const char**)argv);
        my_sim_type::define_parameters(parameters);

        if (parameters.help_requested(std::cout) ||
            parameters.has_missing(std::cout)) {
            return 1;
        }
    
        std::cout << "Creating simulation" << std::endl;
        my_sim_type sim(parameters); 

        // If needed, restore the last checkpoint
        std::string checkpoint_file = parameters["checkpoint"].as<std::string>();
        
        if (parameters.is_restored()) {
            std::cout << "Restoring checkpoint from " << checkpoint_file
                      << std::endl;
            sim.load(checkpoint_file);
        }

        // Run the simulation
        std::cout << "Running simulation" << std::endl;
        sim.run(alps::stop_callback(size_t(parameters["timelimit"])));
\end{lstlisting}

The results of the simulation (the accumulated observables) are
retrieved at line~48 (Listing~\ref{main-cpp-example-collect}). The \verb|results| object is streamed to
standard output which prints out all the accumulated
results. Individual observables can be accessed by name, as shown by
lines~56--62. The observable is represented by an object of type
\verb|result_wrapper| from namespace \verb|alps::accumulators|; each
object has associated mean value, error bar and autocorrelation
length. It is possible to conduct arithmetic operations and compute
functions of the \verb|result_wrapper| objects, as illustrated by
line~65 computing Binder cumulant (per equation~\eqref{eq:ising-binder-cumulant}); the error bars of the operands are
propagated when an expression is evaluated. Finally, at lines~73--76
the simulation parameters and the results are saved to a file in \HDF
format; the name of the file is passed via input parameter
``outputfile'' which is defined earlier in the simulation
implementation file \code{ising.cpp} (by the call at line~15 of
Listing~\ref{ising-cpp-example-defpar}).

\begin{lstlisting}[caption={\textbf{main.cpp} file: Main program, sequential version: Collecting the results},
 label=main-cpp-example-collect, numbers=right, firstline=42, firstnumber=42]

        // Checkpoint the simulation
        std::cout << "Checkpointing simulation to " << checkpoint_file
                  << std::endl;
        sim.save(checkpoint_file);

        alps::results_type<my_sim_type>::type results = alps::collect_results(sim);

        // Print results
        {
            using alps::accumulators::result_wrapper;
            std::cout << "All measured results:" << std::endl;
            std::cout << results << std::endl;
            
            std::cout << "Simulation ran for "
                      << results["Energy"].count()
                      << " steps." << std::endl;

            // Assign individual results to variables.
            const result_wrapper& mag4=results["Magnetization^4"];
            const result_wrapper& mag2=results["Magnetization^2"];

            // Derived result:
            const result_wrapper& binder_cumulant=1-mag4/(3*mag2*mag2);
            std::cout << "Binder cumulant: " << binder_cumulant
                      << " Relative error: "
                      << fabs(binder_cumulant.error<double>()/
                              binder_cumulant.mean<double>())
                      << std::endl;
            
            // Saving to the output file
            std::string output_file = parameters["outputfile"];
            alps::hdf5::archive ar(boost::filesystem::path(output_file), "w");
            ar["/parameters"] << parameters;
            ar["/simulation/results"] << results;
        }
        return 0;
    } catch (const std::runtime_error& exc) {
        std::cout << "Exception caught: " << exc.what() << std::endl;
        return 2;
    } catch (...) {
        std::cout << "Unknown exception caught." << std::endl;
        return 2;
    }
}
\end{lstlisting}

\subsection{Checkpointing}
\label{subsec:checkpointing}
Listing~\ref{main-cpp-example-checkpoint} focuses on the part of the
main program that is relevant for checkpointing. Once the simulation
is completed, at line~46 of the file \code{main.cpp} the state of the
simulation object is saved to a file via a call to method
\verb|save()|, which takes the file name as an argument. If the first
argument to the program is a name of a file in \HDF format, the
parameter object constructor restores the parameters (including their
associated types and values) from the file (the values of individual
parameters still can be overridden from the command
line). Lines~33--37 check if the parameters are read from an
\HDF-formatted file, and, if so, the simulation object is restored
from the checkpoint file via call to method \verb|load()|, which, like
\verb|save()|, takes the file name as an argument. Both \verb|save()|
and \verb|load()| methods are provided by the base simulation class
\verb|alps::mcbase|.

\begin{lstlisting}[caption={\textbf{main.cpp} file: Main program, sequential version: Checkpointing},
 label=main-cpp-example-checkpoint, numbers=right, firstline=30, firstnumber=30, lastline=47]
        // If needed, restore the last checkpoint
        std::string checkpoint_file = parameters["checkpoint"].as<std::string>();
        
        if (parameters.is_restored()) {
            std::cout << "Restoring checkpoint from " << checkpoint_file
                      << std::endl;
            sim.load(checkpoint_file);
        }

        // Run the simulation
        std::cout << "Running simulation" << std::endl;
        sim.run(alps::stop_callback(size_t(parameters["timelimit"])));

        // Checkpoint the simulation
        std::cout << "Checkpointing simulation to " << checkpoint_file
                  << std::endl;
        sim.save(checkpoint_file);
\end{lstlisting}

However, to support checkpointing, user's simulation class
\verb|ising_sim| must provide \verb|save()| and \verb|load()| methods
that take a reference to an object of \verb|alps::hdf5::archive|
type. In our example code, those methods are defined in the simulation
implementation file \verb|ising.cpp|
(Listing~\ref{ising-cpp-example-saveload}), lines~125--136
(\verb|save()|) and~139--155 (\verb|load()|).

\begin{lstlisting}[caption={\textbf{ising.cpp} file: Implementation
  of the simulation class: Checkpointing},
  label=ising-cpp-example-saveload,numbers=right,
 firstline=124, firstnumber=124]
// Saves the state to the hdf5 file
void ising_sim::save(alps::hdf5::archive & ar) const {
    // Most of the save logic is already implemented in the base class
    alps::mcbase::save(ar);
    
    // We just need to add our own internal state
    ar["checkpoint/spins"] << spins;
    ar["checkpoint/sweeps"] << sweeps;
    ar["checkpoint/current_energy"] << current_energy;
    ar["checkpoint/current_magnetization"] << current_magnetization;
    
    // The rest of the internal state is saved as part of the parameters
}

// Loads the state from the hdf5 file
void ising_sim::load(alps::hdf5::archive & ar) {
    // Most of the load logic is already implemented in the base class
    alps::mcbase::load(ar);

    // Restore the internal state that came from parameters
    length = parameters["length"];
    thermalization_sweeps = parameters["thermalization"];
    // Note: `total_sweeps` is not restored here!
    beta = 1. / parameters["temperature"].as<double>();


    // Restore the rest of the state from the hdf5 file
    ar["checkpoint/spins"] >> spins;
    ar["checkpoint/sweeps"] >> sweeps;
    ar["checkpoint/current_energy"] >> current_energy;
    ar["checkpoint/current_magnetization"] >> current_magnetization;
}
\end{lstlisting}

 Most of the logic is implemented in the
corresponding methods of the parent class; the \verb|save()| method
saves only the part of the internal state that changed since
construction of the object (lines~130--133), and the \verb|load()|
method both restores those variables (lines~151--154) and
re-initializes the rest of the internal state from the parameters
(lines~144--147). 

For all basic types and many common \CXX aggregate types (such as STL
vectors, vectors of STL vectors, and others) the saving to and loading
from an \HDF archive are supported via a simple streaming
operators \verb|<<| and \verb|>>|, as shown in lines~131--133
and~152--154. However, to support archiving of the \verb|spins| object
(lines~130 and~151) which is of the user-defined type
\verb|storage_type|, similar methods \verb|save()| and \verb|load()|
must be defined in the corresponding class, as discussed in
subsection~\ref{subsec:impl-details}.

\subsection{Scheduling and parallelization}
\label{subsec:scheduling}

Normally, the sequential simulation terminates once it has run to
completion, as determined by the value returned by the method
\verb|fraction_completed()| of the simulation class. However, the
method \verb|run()| of the simulation base class \verb|alps::mcbase|
accepts a function object argument: the ``stop callback''. This
callback is called repeatedly during the simulation and is expected
to return a boolean value: \verb|false| if the simulation can
continue, \verb|true| if the simulation must be stopped. The library
provides convenience function objects of class
\verb|alps::stop_callback|. The class constructor takes the maximum
simulation time (in seconds) as an argument; the constructed function
object returns \verb|true| if the maximum simulation time has elapsed
since the moment of the object construction, or if the program is
interrupted by a signal. The use of the \verb|alps::stop_callback|
class facilitates a ``graceful'' termination of the simulation in the
case of timeout or interruption by a user.

In order to parallelize the simulation, only minimal changes are
required in the main program, and no changes are needed in the
simulation class. The parallel version of the main program is given by
Listing~\ref{main-mpi-cpp-example}. 

\begin{lstlisting}[caption={\textbf{main\_mpi.cpp} file: Main program, parallel version}, label=main-mpi-cpp-example,numbers=right]
#include "ising.hpp"
#include <iostream>
#include <alps/accumulators.hpp>
#include <alps/mc/api.hpp>
#include <alps/mc/mcbase.hpp>
#include <alps/mc/stop_callback.hpp>
#include <alps/mc/mpiadapter.hpp>

int main(int argc, char* argv[])
{
    // Define the type for the simulation
    typedef alps::mcmpiadapter<ising_sim> my_sim_type;

    // Initialize the MPI environment, and obtain the WORLD communicator
    alps::mpi::environment env(argc, argv);
    alps::mpi::communicator comm;
    const int rank=comm.rank();
    const bool is_master=(rank==0);

    try {
        // Creates the parameters for the simulation
        // If an hdf5 file is supplied, reads the parameters there
        if (is_master) std::cout << "Initializing parameters..." << std::endl;

        // This constructor broadcasts to all processes
        alps::params parameters(argc, (const char**)argv, comm);
        my_sim_type::define_parameters(parameters);

        if (parameters.help_requested(std::cout) ||
            parameters.has_missing(std::cout)) {
            return 1;
        }
    
        std::cout << "Creating simulation on rank " << rank << std::endl;
        my_sim_type sim(parameters, comm); 

        // If needed, restore the last checkpoint
        std::string checkpoint_file = parameters["checkpoint"].as<std::string>();
        if (!is_master) checkpoint_file+="."+boost::lexical_cast<std::string>(rank);
        
        if (parameters.is_restored()) {
            std::cout << "Restoring checkpoint from " << checkpoint_file
                      << " on rank " << rank << std::endl;
            sim.load(checkpoint_file);
        }

        // Run the simulation
        std::cout << "Running simulation on rank " << rank << std::endl;
        sim.run(alps::stop_callback(size_t(parameters["timelimit"])));

        // Checkpoint the simulation
        std::cout << "Checkpointing simulation to " << checkpoint_file
                  << " on rank " << rank << std::endl;
        sim.save(checkpoint_file);

        alps::results_type<my_sim_type>::type results = alps::collect_results(sim);

        // Print results
        if (is_master) {
            using alps::accumulators::result_wrapper;
            std::cout << "All measured results:" << std::endl;
            std::cout << results << std::endl;
            
            std::cout << "Simulation ran for "
                      << results["Energy"].count()
                      << " steps." << std::endl;

            // Assign individual results to variables.
            const result_wrapper& mag4=results["Magnetization^4"];
            const result_wrapper& mag2=results["Magnetization^2"];

            // Derived result:
            const result_wrapper& binder_cumulant=1-mag4/(3*mag2*mag2);
            std::cout << "Binder cumulant: " << binder_cumulant
                      << " Relative error: "
                      << fabs(binder_cumulant.error<double>()/
                              binder_cumulant.mean<double>())
                      << std::endl;
            
            // Saving to the output file
            std::string output_file = parameters["outputfile"];
            alps::hdf5::archive ar(boost::filesystem::path(output_file), "w");
            ar["/parameters"] << parameters;
            ar["/simulation/results"] << results;
        }
        return 0;
    } catch (const std::runtime_error& exc) {
        std::cout << "Exception caught: " << exc.what() << std::endl;
        env.abort(2);
        return 2;
    } catch (...) {
        std::cout << "Unknown exception caught." << std::endl;
        env.abort(2);
        return 2;
    }
}
\end{lstlisting}

At line~12, the
\verb|alps::mcmpiadapter| template parametrized by the user's
simulation class (in our example, \verb|ising_sim|) is used as the
simulation class. The \MPI environment is initialized and an \MPI
communicator is obtained (lines~15--17). The parameters object is
created using a special broadcasting constructor (line~26) that takes
the \MPI communicator as an argument. Likewise, the parallel simulation
constructor takes the communicator as its argument at line~35 (and ensures that each of the parallel simulation clones has its random number generator initialized with a different seed). Line~39
accounts for the fact that each of the parallel \MPI processes now has
its own checkpoint file. The results are seamlessly retrieved at
line~56 from all parallel processes; the master process is responsible
for processing and printing the results, as checked at
line~59. Finally, if an exception is raised during the simulation, all
\MPI processes are aborted (lines~89 and~93). 

In the case of the parallel execution, the simulation is considered
completed if the \emph{sum} of values returned by
\verb|fraction_completed()| methods of all parallel instances of the
simulation class reaches $1$, or if a termination requested by the
``stop callback'' in any of the instances. However, to reduce
unnecessary interprocess communications, the completeness condition is
checked at varying time intervals, within the limits specified by
input parameters ``Tmin'' and ``Tmax''. 
It should be noted that the check of the completeness condition is a collective 
operation: this must be taken into account if the user code invokes \MPI collective
operations on its own. It is therefore recommended, to avoid a possibility of a deadlock,
that any collective operation initiated inside \verb|update()| or \verb|measure()| method
should be completed before the method returns.

\subsection{Other implementation details}
\label{subsec:impl-details}
The 2D-array storage type to keep values of spins is presented in
Listing~\ref{storage-type-hpp-example}. For the sake of clarity and
simplicity, spins are stored as a vector of integer vectors. Note that
the class also defines method for saving to and loading from an \HDF
file (lines~27--33).
\begin{lstlisting}[caption={\textbf{storage\_type.hpp} header file: Implementation
  of a 2D-array}, label=storage-type-hpp-example,numbers=right]
#pragma once

#include <vector>
#include <alps/hdf5.hpp>

// Storage class for 2D spin array.
// Implemented as vector of vectors for simplicity.
class storage_type {
  private:
    std::vector< std::vector<int> > data_;
  public:
    // Constructor
    storage_type(int nrows, int ncols):
      data_(nrows, std::vector<int>(ncols))
    {}

    // Read access
    int operator()(int i, int j) const {
        return data_[i][j];
    }
    // Read/Write access
    int& operator()(int i, int j) {
        return data_[i][j];
    }

    // Custom save
    void save(alps::hdf5::archive& ar) const {
        ar["2Darray"] << data_;
    }
    // Custom load
    void load(alps::hdf5::archive& ar) {
        ar["2Darray"] >> data_;
    }
};
\end{lstlisting}

\subsection{Sample run and results}
For the sake of the example, some parameters for the run are provided via a parameter file 
(see Listing~\ref{ini-file-example}). 
\begin{lstlisting}[caption={\textbf{ising2\_mc.ini} parameter file}, label=ini-file-example,
numbers=right,language=sh]
# Size of the box:
length=4
# Time limit
timelimit = 3600
\end{lstlisting}
\vspace{1.5\baselineskip}
Running of the parallel version on 2 CPU cores with example input parameters produces
the following:
\begin{outputlisting}
$ mpiexec -n 2 ./ising2_mc_mpi ising2_mc.ini --temperature=5 --timelimit=60
Initializing parameters...
Creating simulation on rank 0
Running simulation on rank 0
Creating simulation on rank 1
Running simulation on rank 1
Checkpointing simulation to ising2_mc.clone.h5.1 on rank 1
Checkpointing simulation to ising2_mc.clone.h5 on rank 0
All measured results:
AbsMagnetization: Mean +/-error (tau): 0.342837 +/-0.000101316(12.6489)
Energy: Mean +/-error (tau): -0.456241 +/-0.000168541(10.7416)
Magnetization: Mean +/-error (tau): -0.000150723 +/-0.000275311(31.4836)
Magnetization^2: Mean +/-error (tau): 0.175277 +/-9.02575e-05(13.7491)
Magnetization^4: Mean +/-error (tau): 0.0730068 +/-6.40981e-05(12.6128)

Simulation ran for 147922660 steps.
Binder cumulant: Mean +/-error (tau): 0.207852 +/-0.000218572(12.6128) Relative error: 0.00105158
\end{outputlisting} 
and creates files \verb|ising2_mc.out.h5|,
\verb|ising2_mc.clone.h5| and \verb|ising2_mc.clone.h5.1| in the current directory.

\begin{figure}[ht]
  \centering
  \begin{tabular}{ll}
  (a)\raisebox{-\height+3ex}[3ex][\height-3ex]{\includegraphics[origin=c,width=0.5\textwidth]{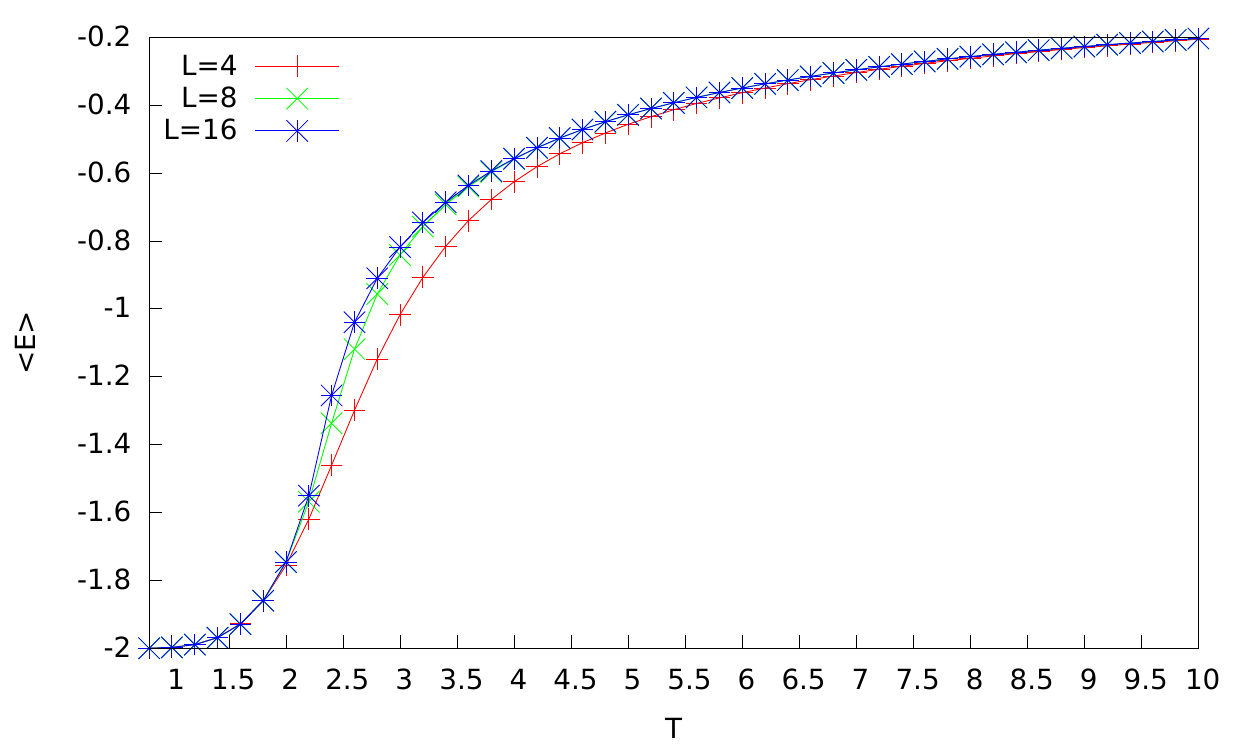}}
  &
  (b)\raisebox{-\height+3ex}[3ex][\height-3ex]{\includegraphics[width=0.5\textwidth]{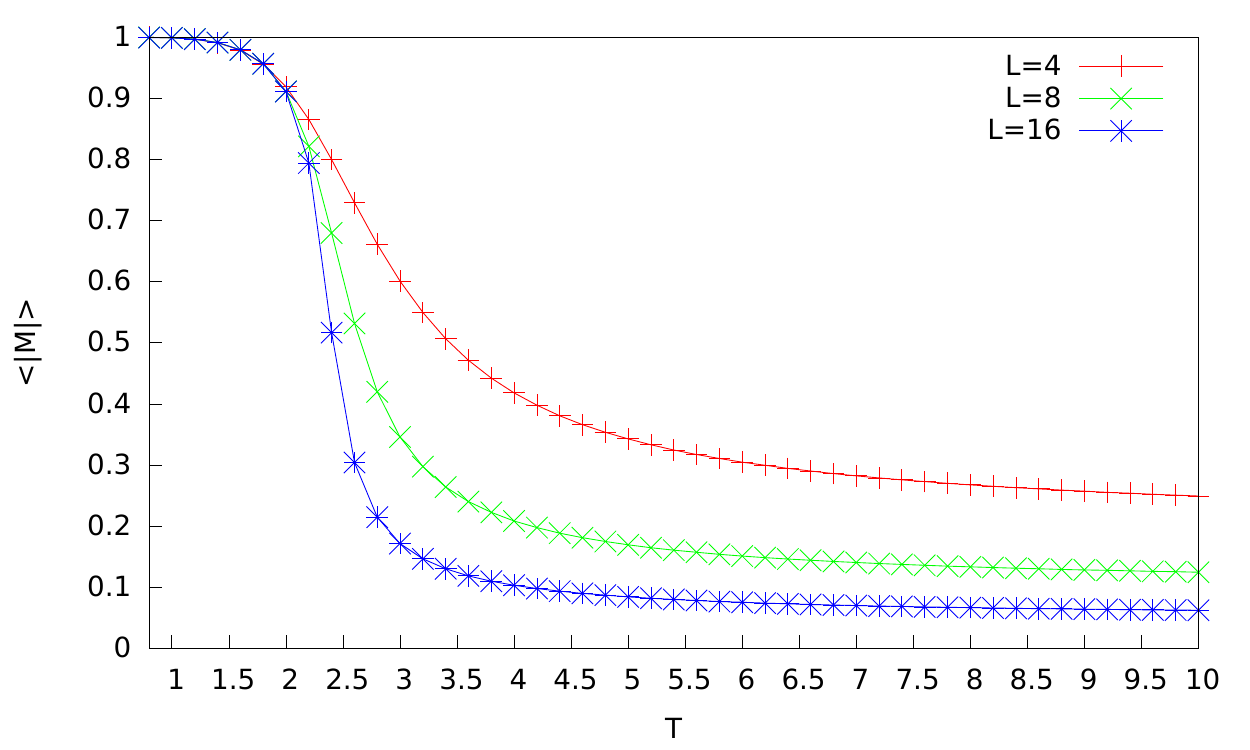}}
  \\
  (c)\raisebox{-\height+3ex}[3ex][\height-3ex]{\includegraphics[width=0.5\textwidth]{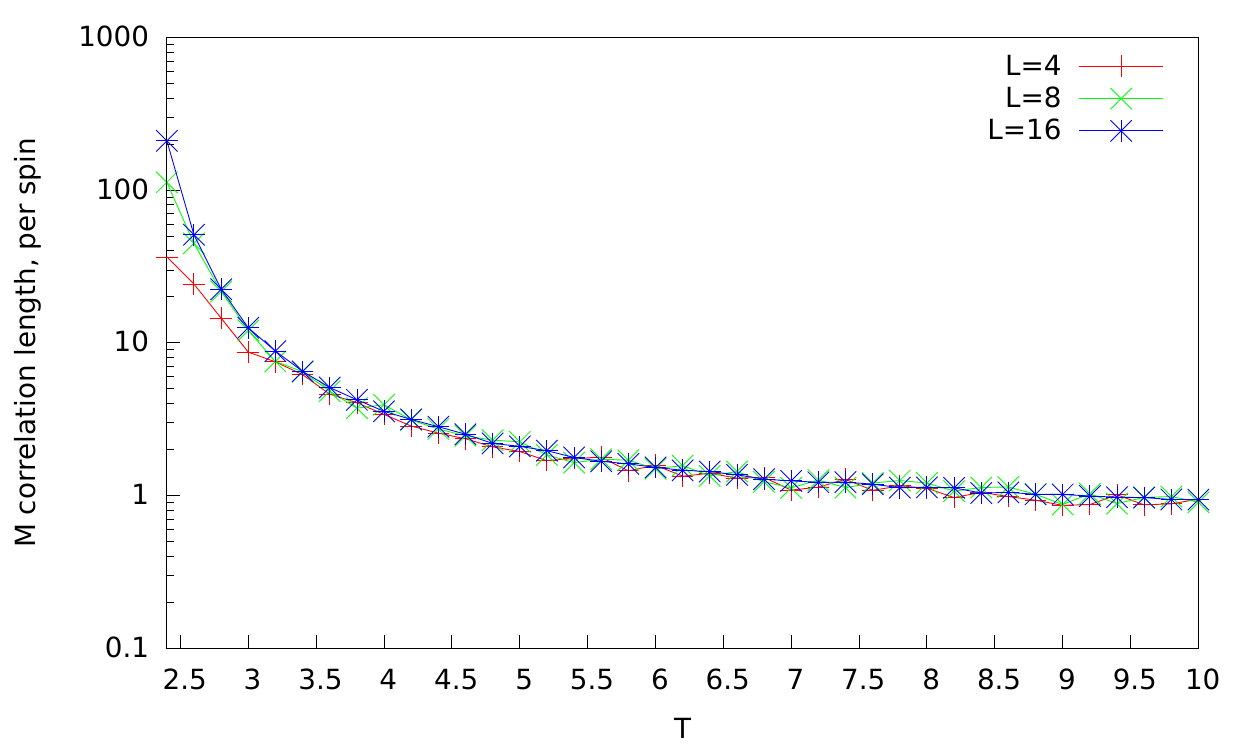}}
  &
  (d)\raisebox{-\height+3ex}[3ex][\height-3ex]{\includegraphics[width=0.5\textwidth]{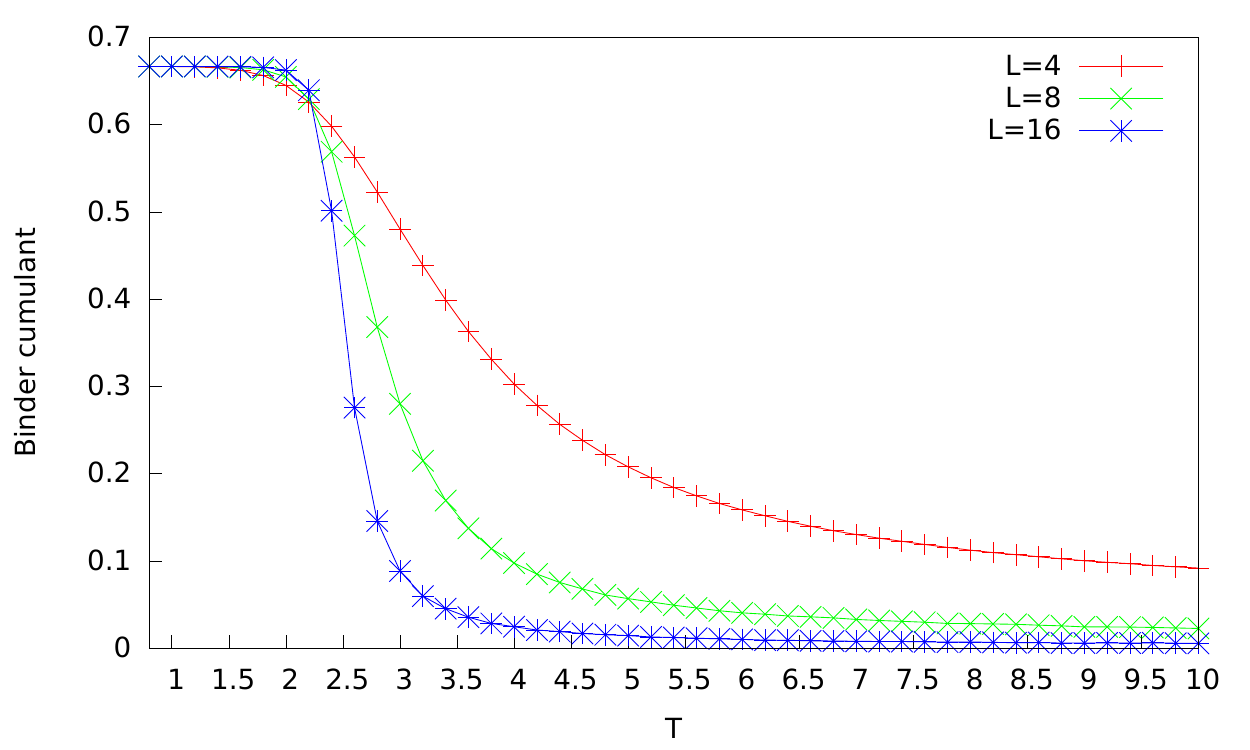}}
  \end{tabular}
  \caption{The results of the 2D Ising model calculation implemented using
    ALPS core libraries, with periodic box of sizes $L=4$, $L=8$ and $L=16$. The
    quantities plotted against the temperature T are: 
    (a)~Mean energy $\langle E\rangle$ per spin;
    (b)~Absolute value of magnetization $\langle|M|\rangle$ per spin;
    (c)~Autocorrelation length, per spin (logarithmic scale) of magnetization $M$;
    (d)~Binder cumulant for magnetization distribution.}
  \label{fig:results}
\end{figure}

The program was run on a set of pre-generated parameter files with
varying temperature and system size; the standard output was parsed
and formatted for plotting by a simple script. The results are
presented in figure~\ref{fig:results}.

\section{Example: Green's Functions interface}
\label{sec:example-gf}

Listing~\ref{gf-example} illustrates working with Green's functions.
\begin{lstlisting}[caption={Working with Green's Function demo}, 
label=gf-example,numbers=right]
#include <alps/gf/gf.hpp>
#include <alps/gf/tail.hpp>

#include <boost/array.hpp>

namespace g=alps::gf;

// Generates points for momentum mesh
g::momentum_index_mesh::container_type generate_momentum_mesh();

void Demo() {
    const int nspins=2;
    const int nfreq=10;
    const double beta=5;

    // Construct the meshes:
    g::matsubara_positive_mesh m_mesh(beta, nfreq);
    g::momentum_index_mesh k_mesh(generate_momentum_mesh());
    g::index_mesh s_mesh(nspins);
    
    // construct a GF using a pre-defined convenience type
    g::omega_k_sigma_gf gf(m_mesh, k_mesh, s_mesh);
    // initialize a GF to all-zeros
    gf.initialize();

    // Make indices:
    g::matsubara_index omega;
    omega=4;
    g::momentum_index ii(2);
    g::index sigma(0);

    // Assign a GF element:
    gf(omega,ii,sigma)=std::complex<double>(3,4);

    // Density matrix as a double-valued GF
    // on momentum and integer-index space:
    typedef
        g::two_index_gf<double, g::momentum_index_mesh, g::index_mesh>
        density_matrix_type;
    
    // Construct the object:
    density_matrix_type denmat=density_matrix_type(k_mesh,s_mesh);

    // prepare diagonal matrix
    const double U=3.0;
    denmat.initialize();
    // loop over first mesh index:
    for (g::momentum_index i=g::momentum_index(0);
         i<denmat.mesh1().extent(); ++i) {
        denmat(i,g::index(0))=0.5*U;
        denmat(i,g::index(1))=0.5*U;
    }

    // construct a tailed GF using predefined convenience type:
    g::omega_k_sigma_gf_with_tail gft(gf); 
    gft.set_tail(0, denmat);              // set the tail

    density_matrix_type gftail=gft.tail(0); // retrieve the tail

    // access the tailed GF element
    std::complex<double> x=gft(omega,ii,sigma);
}

/// Generates 4 2-D points k_1,...,k_4 where each point k=(k_x,k_y)
g::momentum_index_mesh::container_type generate_momentum_mesh() {
    const boost::array<int,2> dimensions={{4,2}};
    g::momentum_index_mesh::container_type mesh_points(dimensions);
    mesh_points[0][0]=0;    mesh_points[0][1]=0;    // (0,0)
    mesh_points[1][0]=M_PI; mesh_points[1][1]=M_PI; // (pi,pi)
    mesh_points[2][0]=M_PI; mesh_points[2][1]=0;    // (pi,0)
    mesh_points[3][0]=0;    mesh_points[3][1]=M_PI; // (0,pi)
    return mesh_points;
}
\end{lstlisting}
At lines~17--19 three meshes are constructed: a Matsubara mesh, a mesh
in momentum space, and a general index mesh used here for the spin
variable. Then, at line~22 a predefined convenience type is used to
create a Green's function defined on a Cartesian product of these
three meshes. An element of the Green's function can only be addressed
by a tuple of 3 indices each belonging to a corresponding mesh, as
shown in the listing: lines~27--30 define the indices and assign
values to them, and line~33 accesses the Green's function element.

Lines~37--39 define a Green's function type \verb|density_matrix_type|
explicitly as a floating-point (\CXX \verb|double|) valued function
defined on a Cartesian product of two meshes: the momentum mesh and
the spin mesh. Here, a density matrix represented as an
object of this type. The loop over the momentum index (lines~48--52)
illustrates the use of inequality and increment operators on a mesh
index to set values of the density matrix. 

At line~55 a Green's function with a known tail is created from a
regular Green's function. The tail encapsulates a high frequency
expansion of a Green's function as:

\begin{equation}
  \label{eq:tailed-gf}
  G(i\omega_{n},x\ldots) = c_{0}(x\ldots) 
  + \frac{c_{1}(x\ldots)}{i\omega_{n}} 
  + \frac{c_{2}(x\ldots)}{(i\omega_{n})^{2}}
  + \cdots
\end{equation}
where $\omega_{n}$ is Matsubara frequency, $c_{j} (j=0,\ldots)$ are
tail coefficients of $j$-th order, and $x\ldots$ represent all other
arguments of a multi-dimensional Green's function. Note that the tail
coefficients in this representation are themselves functions defined
over $(x\ldots)$.

At line~56 of the listing~\ref{gf-example}, the density matrix
\verb|denmat| is set as the 0-th order tail coefficient.

Finally, lines~65--73 define the function used to generate the momentum
space mesh: the mesh contains 4 points, and each point $(k_{x},k_{y})$
lies in a 2-dimensional momentum space.

\end{document}